\documentclass[preprint2]{aastex}
\usepackage{latexsym,epsfig}
\usepackage{aastexug}   
\newcommand{\etal}{et al. }
\begin{document}

\title{A CCD TIME-SERIES PHOTOMETER}

\author{R. E. Nather and Anjum S. Mukadam}

\affil{Department of Astronomy, University of Texas at Austin, Austin, TX -78712, U.S.A.}
\email{nather@astro.as.utexas.edu; anjum@astro.as.utexas.edu}

\begin{abstract}

We describe a high speed time-series CCD photometer for the prime focus of the
82-in (2.1\,m) telescope at McDonald Observatory, and summarize the
observational results we have obtained since it was placed into
regular use in February, 2002. 
We compare this instrument with the
three-channel time-series photometers we have previously used in
the asteroseismological study of pulsating white dwarf stars, which
used photomultiplier tubes (PMT) as the detectors. 
We find the CCD
instrument is about 9 times more sensitive than the PMT instruments
used on the same telescope for the same exposure time. 
We can therefore find and measure variable
white dwarf stars some 2.4 magnitudes fainter than before, significantly
increasing the number of such objects available for study.

\end{abstract}


\section{Introduction}

We have designed, and placed into operation, a CCD camera system
optimized for high speed time-series measurements of oscillating white dwarf
stars. Our experience indicates that CCD instruments designed for
more general use will have characteristics that are unacceptable
for rapid time-series measurements, or seriously compromise the 
data quality when used for this purpose. Until this instrument was 
available, we relied on time-series photometers using photomultiplier 
(PMT) detectors for these measurements (Nather \& Warner 1971; Kleinman,
Nather, \& Phillips 1996).

We are now in a position
to compare and contrast these two approaches to the same measurements,
and to demonstrate where CCDs are superior for this work and where they 
are not.

One basic goal in the design of this instrument was to take advantage
of the improved quantum efficiency that CCD detectors offer, so that
we could obtain usable data on fainter stars than the PMT instruments
could measure, and data of better quality on those they could. We
also hoped to discover more variable white dwarf stars than were
already known, to increase the limited number of objects available
for asteroseismological study. A more long-term goal was to provide
an instrument whose timing accuracy was high enough to allow a search
for small deviations from the smooth secular change in pulsation 
frequency due to white dwarf cooling (e.g. Kepler \etal 2000), opening the possibility that such 
deviations, if found to be periodic, could demonstrate the presence of 
planet-sized objects in orbit around the white dwarf star (Winget \etal 2003;
Mullally \etal 2003).

We have achieved our short-term goals in the 15 months the instrument
has been in operation, and have established a list of target objects
for the longer-term search for primordial planets. Our search for white 
dwarf variables has already more than doubled the number available for 
study (Mukadam \etal 2003a), due primarily to increased instrumental sensitivity. Our practical 
limit using the PMT photometers on the 82-in telescope at McDonald 
Observatory was about magnitude 17.0; we now obtain light curves of 
comparable quality at magnitude 19.4, a gain of about a factor of 9
in overall sensitivity. To obtain a similar gain with the PMT photometers 
they would have to be attached to a telescope 6\,m in aperture.
 
\section{HARDWARE}

\subsection{The CCD Camera}

Our CCD time-series photometer, which we call Argos, is based on a
commercial CCD camera made by Roper Scientific, the Princeton 
Micromax 512\,BFT~NTE-CCD
\footnote{http://www.roperscientific.com/micromax.html}. Its specifications 
are shown in Table 1.

\begin{deluxetable}{ll}
\tablecolumns{2}
\tablewidth{0pc}
\tablecaption{Summary of Camera Specifications}
\tablehead{}
\startdata
Pixel size:   &         13$\mu $\,$\times $13\,$\mu $\\
Pixel array size: &     512$\times $512, back illuminated\\
On--chip storage:    &     512$\times $512, frame transfer operation\\
Frame transfer time: &  310\,$\mu $\,s\\
Readout rate:        &  1 MHz, 16 bit A/D conversion\\
Readout time:        &  0.28\,s, full frame with no binning\\
Cooling:             &  Thermoelectric + fan air exhaust\\
Chip temperature:    &  -45\,C\\
Readout noise:       &  8 electrons RMS\\
Gain		     &  2 electrons/ADU \\
Dark noise:          &  1--2 ADU/s/pixel\\
Optical coating:     &  broadband anti-reflection\\
Quantum Efficiency:  &  30\% at 3500A, 80\% 4500-6500A, 40\% at 9000A\\
Linearity:	     &  $\sim $ 1\% below 40,000 ADU (saturation at 65,000 ADU)\\
\enddata
\end{deluxetable}

The CCD chip is back-illuminated to improve its blue sensitivity
(most white dwarf stars are blue) and can transfer its 512$\times $512 
pixel image
to the on-chip buffer in 310\,$\mu $\,s.  We have built a mount
to support the camera at the prime focus of the 82-in telescope (F/3.9),
and obtain there an image scale to match the 13\,$\mu \times 13\,\mu $ 
pixel size: 
3.05 pixels per arcsecond. The field of view for this image scale is 
2.8 arcmin on a side, large enough so we have not had any trouble finding 
our targets, along with suitable comparison stars. Since the target can
be placed almost anywhere in the chip, the usable area to search for
comparison stars is about 25 square arcminutes.

The camera incorporates a thermoelectric cooling system that keeps 
the chip at -45\,C, where the dark count of 1--2 ADU/s
is smaller than the counts coming 
from the moonless sky (ca. 3--7 ADU/s). 
The readout noise of 8 electrons RMS is negligible 
for all except the shortest exposure times, where it is comparable to sky 
noise. The image readout time of 280\,ms
is comfortably shorter than our minimum 
exposure time of 1\,s. 
The prime focus mount design (see Figure 1)
includes a manual two-position filter slide, which can double as a dark 
slide when we want to take dark or bias frames. The camera came with 
an internal shutter for this purpose, but we removed it when it proved 
to be unreliable.

\begin{figure}[ht]
\figurenum{1}
\plotone{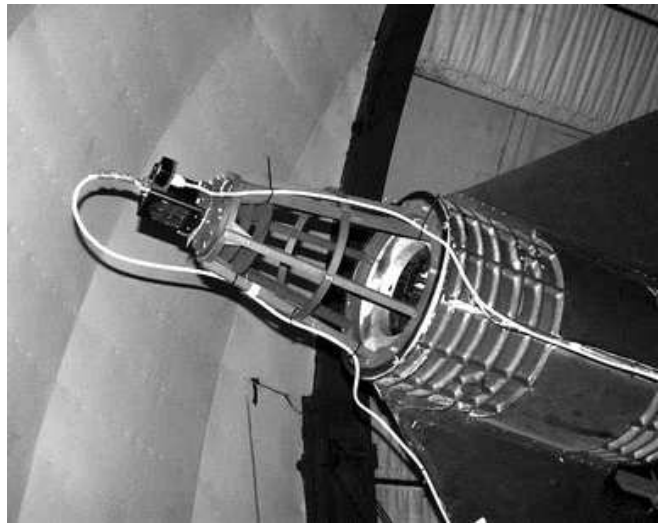}
\figcaption{The CCD photometer Argos on the prime focus of 
the 82-in telescope at McDonald Observatory. Photo by David Doss.}
\end{figure}

The CCD camera connects to the ST-133 controller (electronics box) via a 10 ft.
analog cable, both of which are mounted at the prime focus of the 82-in 
telescope. 
The interface card housed in the camera control computer (see section 2.7)
connects to the controller via a 75 ft. digital high speed communication cable.
Frame transfer operation is started by a single synchronizing pulse
from a timer card (see section 2.6) which serves to end one exposure and
begin the next one;
the pulse is sent to the camera via a 75 ft. co-axial cable. 

We have also attached a smaller uncooled CCD camera (Electrim EDC2000n
\footnote{http://www.electrim.com})
with a wide angle lens to capture regular images of the dome slit 
(see section 2.5).

\subsection{Wavelength Response}

While the CCD chip is more sensitive than the bi-alkali PMT detectors,
its wavelength response is enough different so that combining 
CCD and PMT data on the same star may not be done directly.
Our chief interest lies in blue pulsating DA white dwarfs,
whose pulsation amplitudes are a function
of wavelength (Robinson \etal 1995, Nitta \etal 1998, Nitta \etal 2000).
Including photons redward of the PMT cutoff (ca. 650 nm), which are
less modulated by the pulsation process, reduces the measured
amplitude (Kanaan \etal 2000). We thought we might improve the
signal--to--noise ratio in our light curves by inserting a
blue filter (1mm Schott BG40), and indeed found the measured amplitudes
are higher by about 15\%.
However, photon losses in the filter increased the measurement noise,
and we find no significant improvement in the signal--to--noise
ratio by using it.

\subsection{The Prime Focus Mount}

The camera mounting plate can be moved by x-y adjustment screws over a 
distance of half an inch to center the CCD chip on the optical axis of 
the telescope.  The tip/tilt of the camera can be controlled by a 
push-pull arrangement of screws, to set the CCD chip
perpendicular to the axis. 
The alignment procedure is adequate but 
awkward. 
When properly aligned, the corners of the chip are 2 arcmin 
from the optical axis, where aberration from coma due to the parabolic 
primary mirror is calculated to expand a point image to about 1 arcsecond 
in diameter. We rarely experience sub-arcsecond seeing at McDonald 
Observatory, so we have not been able to verify this calculation.

\subsection{Baffling Argos}
Scattered light was initially a significant problem. 
The first three nights of the commissioning
run proved beyond doubt that the shiny aluminium surfaces 
of the mount had to be darkened; we chose hard black anodizing for
the purpose, as it does not corrode easily.

The single crude baffle in
the original design was replaced with a five-stage baffle
system consisting of two thin plates very close to the camera,
and three other baffles in the body of the mount. The two camera baffles 
and the mount baffle closest to the camera have
square shaped apertures with rounded corners and are derived by projecting the 
light beam backwards from the CCD chip. These openings
are a few percent larger than the converging light beam from the primary.
The other two mount baffles have circular apertures, which are 5--7\%
bigger than the light beam. The edges of all the light baffles 
are at an angle of $45^{\circ }$ with respect to the optic axis to reflect 
light away from the CCD camera.

The original flat-field images were very strange, 
but have now been improved so that people no longer laugh at them. 
Our current images are
flat to within a few percent;
the variation in the flat fields comes from structural 
non-uniformities in the CCD chip itself.
This pattern is stable and can be removed, giving us residual variations 
less than one percent.

\subsection{Collisional Danger to Argos}
One significant problem remains: in its normal position at prime focus,
the camera, and its mount, can collide with structures inside the dome.
The dome slit has a heavy steel bridge spanning its width, and on each 
end of the bridge are hand-cranked pods (called pulpits) that can hold 
an observer; by moving the bridge and cranking a pulpit, an intrepid 
observer can reach the prime focus for visual guiding. This is clearly 
dangerous, and has not seen use since the prime focus (photographic plate) 
camera was retired about 35 years ago, but is still used to help balance 
the telescope and to support a crane that can handle the primary mirror 
for aluminizing. Since we can't remove the bridge and pulpits we must
learn to live with them.

To this end we have added Cyclops, a small uncooled CCD camera 
with a wide-angle
lens ($150^{\circ }$ FOV) to the Argos camera, looking out past the dome
slit at the night sky beyond. With suitable exposure times it can see
the moonless sky as significantly brighter than the inside of the dome,
and can thus define the position of the prime focus mount with respect
to the slit as well as the bridge and pulpit structure (Figure 2). 
The circle shows the 
size and location of the 82-in telescope beam. The edge of the lower
windscreen, just above the bridge/pulpit structure, appears at the lower
left in the image.

\begin{figure}[ht]
\figurenum{2}
\plotone{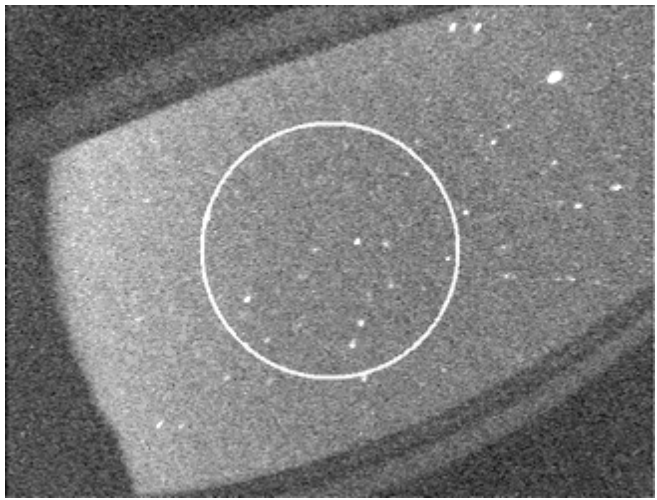}
\figcaption{View from the dome monitoring camera Cyclops: The circle shows
the location and size of the light beam incident on the primary mirror of the 82-in telescope.}
\end{figure}

The camera's dynamic range is large enough to define the prime focus 
mount position in the daytime as well, when the dome is closed and 
illuminated from the inside. We use a dedicated PC to run the camera 
whenever Argos is mounted on the telescope. Suitable software to warn of 
impending disaster, and perhaps prevent it, is now under development.  
The images by themselves are already useful, telling the observer when 
it is time to move the dome. Cyclops has detected raindrops as well.

\subsection{The Timing System}

A time-series photometer must know precisely when an exposure is started,
and precisely how long it took. These are two different requirements,
involving both a time epoch and an interval. The Argos timing system
is based on a GPS clock designed primarily for precision timekeeping,
although position information is also available. It consists of an
oven-controlled crystal oscillator disciplined by signals obtained
from a GPS receiver\footnote{http://www.trimble.com/thunderbolt.html}. The 
time epoch is claimed to have an error of about 50\,ns, considerably 
more precise than we need, but comforting.

We have assembled a simple count-down register on a small circuit card
that can accept the 1 Hz timing pulses from the GPS clock, and can
provide an output pulse to initiate frame transfer in the CCD camera.
The exposure intervals are thus contiguous, and determined directly
from the clocking hardware. The timer card plugs into the parallel
port on the camera control PC, so the countdown value (i.e. the
exposure time) can be set into it from software. Thereafter it
operates independently, initiating frame transfer operations at the
established timing intervals. Exposure times can be set to any integral
number of seconds from 1 to 30.

Immediately following a 1 Hz timing pulse, the GPS clock provides
information from which the precise epoch of the pulse can be determined.
The information arrives encoded in packet form at the serial port of
the PC at 9600 baud, where it can be read and unencoded by the software
control program. The PC serial port is buffered so that the epoch (the 
time and date) for each pulse can be determined even if a second packet
arrives before the first one is read by the program.

A second clock, somewhat less accurate, is available if the camera
control PC is running the Network Time Protocol (NTP\footnote{http://www.ntp.org}) software, which
obtains timing information over the internet by periodically contacting
time servers and adjusting the PC system clock accordingly. It allows
for internet time delays as well as it can, and averages the best
readings it finds to keep the system clock in proper synchronization.
The Argos control program relies primarily on the GPS clock for timing,
but can use the NTP-disciplined system clock if the GPS time signals
are not available. For observer assurance, it compares the time ticks
from the two clocks, and displays their time difference in a status
display window. After both clocks have been running for a few hours,
the time difference is usually within a few milliseconds.

\subsection{The Camera Control Computer}

The PC that controls the camera has fairly modest requirements by
modern standards: it must have a PCI bus to accept the camera control
card, a parallel port (for the timer card), a serial port (for the
GPS packet information), enough memory to run the Linux operating
system comfortably (256 MB is enough, but more is always better) and
enough disk space to hold the images as they arrive (527 Kb each).
Our current camera control PC runs a Pentium III at 1 GHz and is not 
pushed for time.  The software prefers a display resolution of 1280x1024 
so the various windows do not overlap each other. A 17-in LCD works fine.
The PC also needs an ethernet card to connect to the internet, so the
NTP software can discipline the system clock, and to receive pointing
information from the computer that controls the 82-in telescope.
This connection is also used to transfer image data to our Argos
data archive in Austin (slowly), and to allow a remote login to run
the camera for testing purposes (even more slowly).

We usually operate a second PC as well, with access to the disk on
the camera control computer, so arriving image data can be examined
by software not concerned with the data acquisition and recording
process. We also make a more durable copy of the data on CD-ROM.
Someday, when the DVD format wars are over, we may move to that
medium to minimize the number of disks required.

\section {SOFTWARE}

\subsection{The User's View}

\subsubsection{Program Requirements}

The control program is called Quilt 11 (q11), the most recent in a series of
programs designed to control time-series photometers.  The name was
originally chosen because the first of the series, Quilt 1, started as
a patchwork of software routines. It was written in 1970, and has
undergone 10 complete rewrites, in different languages for different
computers, since that time. The basic operations haven't really
changed much.

Once the camera has been set to operate in frame transfer mode,
images arrive via direct memory access (DMA) at the end of each
exposure and appear magically in memory.  The program must first
associate each image with its epoch (start time and date) before
it is written to disk.  This is not quite as straightforward as it
sounds: a new exposure starts when a frame transfer operation
finishes, so the epoch is available right away, but obviously
the image is not --- it's still being exposed.  It only shows up
after the next timing pulse (and its epoch) arrives,
and then only after the readout process has finished.  The
program must keep all this straight so the proper epoch is
associated with the appropriate image.

{\onecolumn
\begin{figure}[ht]
\figurenum{3}
\plotone{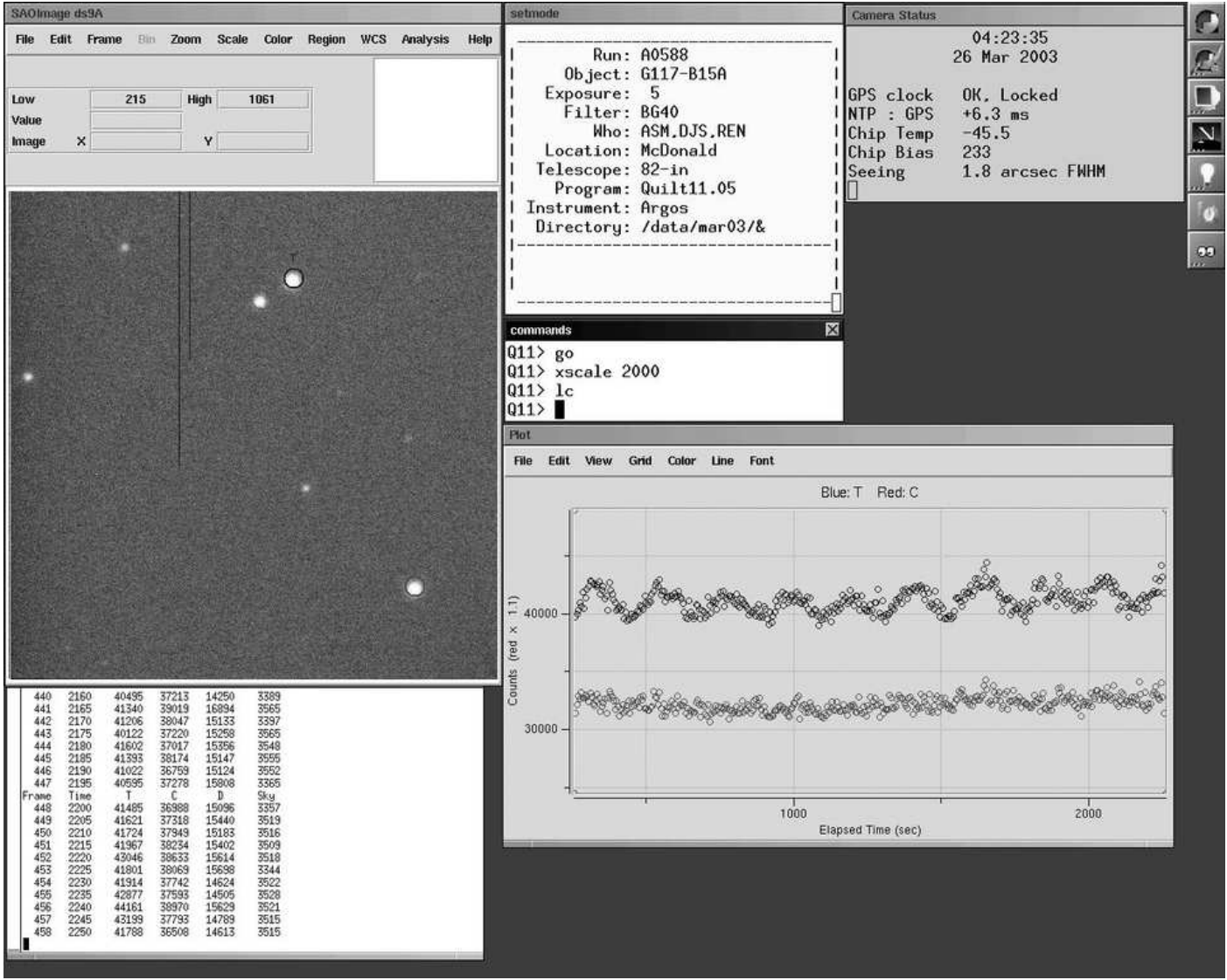}
\figcaption{User's view of Quilt 11, the data acquisition program.}
\end{figure}
}
\twocolumn

The data images are recorded in FITS  
\footnote{http://fits.gsfc.nasa.gov/fits\_intro.html} format, with the 
epoch and other operating parameters in the header. Each image has its 
own file, so the file name must be generated automatically and the 
names must be sequential so they can be kept in proper time order.

\subsubsection{Controlling q11}
Figure 3 shows what the computer ``desktop'' looks like with the program
in operation.
The window labeled ``Commands'' accepts simple typed
keyboard commands (go, stop, abort, etc.) to control the camera, as
well as commands to edit entries in the window labeled ``Setmode''
which provides the q11 program with information and parameters
that it cannot determine for itself.  The user may then
mark the target and comparison stars on the image (small, labelled
circles appear in response to mouse clicks) to enable on-line 
extraction and display of plotted light curves.

In our design the images arrive whenever a frame
transfer pulse is
generated by the clock -- that is, all the
time. These images are shown on the display screen by
a program called DS9, written by
William Joy and his colleagues at SAO and made available as Open Source 
Software.  Plotted light curves are displayed by the DS9 plotting
widget.

Once data recording starts
(in response to the ``go'' command), the observer must act as an aide to 
the program to do things it cannot do for itself: keep the star images on
the chip, keep the dome out of the light path, and be prepared to
shut things down if it rains. In addition to plotting the light
curve, the program also prints columns of numbers on the terminal
used to start the program: the image number, the elapsed time, and
the extracted brightness for each of the marked stars with sky
removed. The last column shows the average sky value that was
subtracted from the target star.

\subsubsection{Simulation}

The program can also be run in simulation mode, without a telescope,
camera or GPS timing system. Previously recorded data images are read
from disk into the same buffer used by the camera, and the timing is 
simulated from the system clock ticks. Users can use this mode to review 
data previously recorded, and to learn how the program works.  The same
code is used in both modes, with very few exceptions. If the program
is started with a command-line pathname to a data run, simulation mode
is assumed; without one, it assumes it must run the camera for real.
Simulation mode goes through all of the motions involved in a real
run but does not write anything to disk.

\subsubsection{Software Design}

The q11 program is written in the C language, and consists of 5 separate
executable processes in simultaneous execution.  
It is designed to run on a PC under the Linux operating system and to 
tolerate the presence of other programs running at the same time on the same 
CPU.  Both incoming time and image data are buffered to maintain proper 
real-time operation.  Details of its architecture and the on-line extraction
algorithms are presented in Appendix A.

\subsubsection{Display}

The data acquisition and on-line extraction process runs in a thread 
separate from the display process because of timing considerations: the 
display routines (DS9 and its plotting widget) are written in Tcl/Tk,
an interpreted language, and are therefore slow. At the shortest
exposure times the acquisition process can easily keep up but the
display routines cannot. Arranged as separate threads of execution,
they run in parallel, so acquisition gets its needed amount of CPU time
even if display falls behind. 
Should this happen, the user still sees
the printed luminosity values appear right away, but the plotted values 
appear in clumps, rather than one at a time, whenever the plot widget 
gets updated.  No data points are lost.
Display of some of the incoming images may be skipped, but the 
DS9 window always shows the most recent one when it is updated.
The q11 program can run under any window manager, but users notice how
much more slowly the display windows are updated using Gnome or KDE,
compared with WindowMaker, which is much smaller and faster.

\section{COMPARING: CCD vs PMT}

\subsection{Good Things}

\subsubsection{Digital Image Preservation}

The most notable improvement offered by the Argos instrument is the
ability to record individual images for each integration for later
inspection and analysis. This is very much like the historical
transition that took place as photographic observations replaced
visual ones. Extracting the measured brightness of the target,
comparison star and sky is done directly with the 3-channel PMT 
photometer, in the equivalent of three large pixels, through
fixed apertures that isolate them from the rest of the stars in
the field. What they see is what you get, with no going back.

The digital images arriving from the CCD are recorded as individual
disk files and can be replayed without loss in the same time
sequence as they were taken, with their time intervals the same as
the original exposure times for simulation, or faster for analysis
and reduction. Different extraction techniques can be applied for 
comparison, and the light curves with the highest signal-to-noise
ratio chosen for further analysis. 
When a data point appears that
does not fit well with those surrounding it, the corresponding image
can be examined, and often reveals the cause (``Oh.  I guess that's
where I dropped my flashlight.'') We find far more satellite tracks
through the images than we ever expected.

\subsubsection{Greater sensitivity}

\begin{figure}[ht]
\figurenum{4}
\plotone{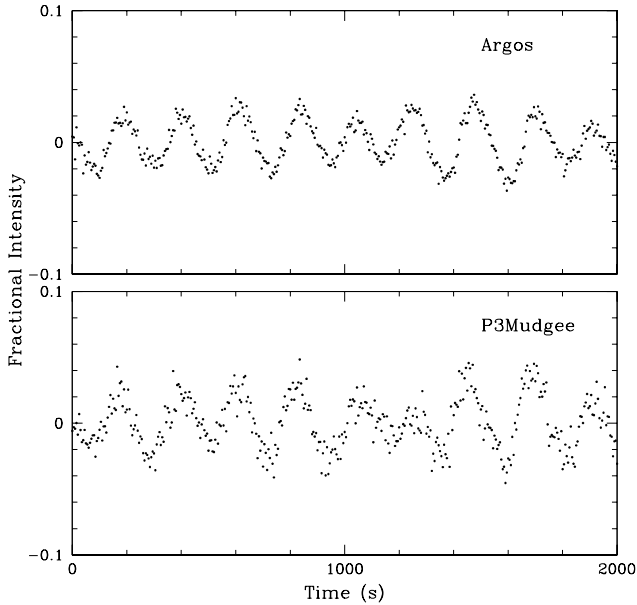}
\figcaption{The top panel shows the light curve of G\,117-B15A taken with Argos at an
exposure time of 5\,s on 12 November, 2001. The lower panel shows the light curve of the
same star observed with the 3 channel PMT photometer, P3Mudgee, at the same telescope, and
with the same exposure time on 20 December, 2001.}
\end{figure}

The higher quantum efficiency of the CCD coupled to the greater
bandwidth gives higher photon counts from the same target
stars, compared to the PMT photometers. 
Argos is designed such that light from the primary mirror
forms an image of the star field directly on the CCD, so it
has fewer optical surfaces than the PMT photometers.
Figure 4 shows the light curves of the same white dwarf pulsator 
G\,117--B15A taken
with the two different instruments, 5 weeks apart, on the 82-in 
telescope with 5\,s exposures.
Observing conditions were excellent for the PMT run, but only
average for the CCD; even so, the CCD data are less noisy.

The ability to measure fainter stars is illustrated in Figure 5.  The
target star, a new DA variable WD0815+4437 (Mukadam \etal 2003a),
has a B magnitude comparable to 19.3, and clearly shows the pulsations;
the FT shows two significant peaks (probably unresolved in this short run)
and hints at more. Smoothing the light curve with a running average
of 3 data points (to suppress noise at higher frequencies) shows
the variations more clearly.

\begin{figure}[ht]
\figurenum{5}
\plotone{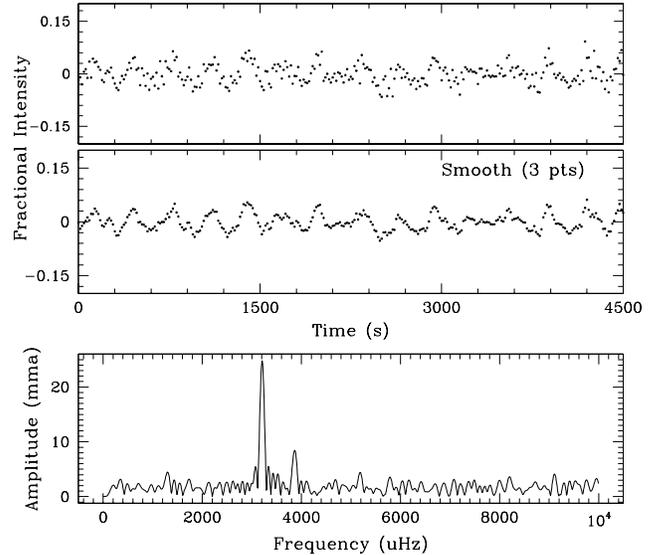}
\figcaption{The top panel shows the light curve of WD0815+4437, a faint (B$\approx $19.3) new DA
variable (Mukadam \etal 2003a). We acquired the data on 1 February, 2002,
with an exposure time of 15\,s including the BG40 filter.
The middle panel shows the same light curve
after a 3 point smoothing. The bottom panel shows the FT of the light curve,
where the yscale is in mma ($1\,mma = 0.1\% \Delta I/I$).}
\end{figure}

\subsubsection{Marginal Photometric Conditions}

The PMT photometers separate the target and comparison star fields
optically, and do this before the image of the field is formed
at the focal plane. This works, but introduces a vignetting property
which the users must be aware of, and avoid. This means the comparison
star must always be at least 3 arcminutes away. Changing transparency
from thin cloud does not always act on the target and comparison stars
at the same time, so removing the effects of cloud by dividing the
target light curve with that of the comparison star does not always
work very well.

This technique of cloud removal works far better on the CCD images.
In Figure 6 we show our data on the new DA variable WD0949-0000 
(Mukadam \etal 2003a, 2003b)
taken through light cloud with 10\,s exposures on 2 April, 2003.
The top panel shows the sum of 2 comparison stars,
collectively brighter than the target by a factor of 150.
The center panel shows the faint (B$\approx $18.8)
target star. The cloud-induced variations disappear when
the target is divided by the comparison light curve.

\begin{figure}[ht]
\figurenum{6}
\plotone{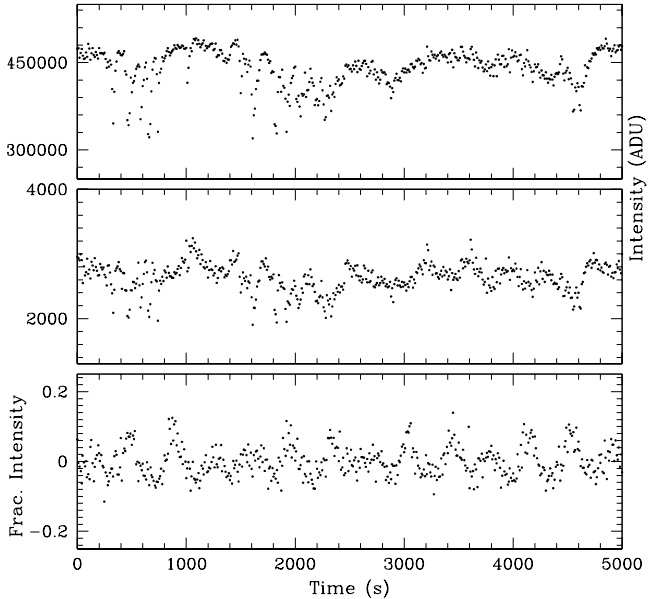}
\figcaption{Cloudy Weather conditions: 
The top panel shows the summed light curve of two comparison stars in the field;
the center panel shows the raw light curve of the (much fainter)
target star, a new DA variable.
The bottom panel shows the reduced data on the star
after it has been divided by the comparison light curve.} 
\end{figure}

\subsubsection{The Human Factor}

The PMT photometers require considerable skill and experience on the
observer's part to get everything set up properly, and to tend the
observations continually during a run. Failing to align the two
stars properly in their apertures, or to guide often enough but not
too often, takes some time to learn. Inexperienced observers
can (and do) take data of poor quality until they have made most of
the common mistakes and learned from them. The Argos photometer
demands far less skill and experience; once the CCD instrument is set
up properly, far less is demanded of an observer, and inexperienced
observers can (and do) take good quality data on their first observing
run with the instrument.  Target starfields are much easier to locate
and verify, and, unlike the PMT photometers, faint targets are as
easy to find as bright ones.

\subsection{Bad Things}

\subsubsection{Read noise}

A photomultiplier detector amplifies individual photoelectrons until
they can be readily detected as individual events; a CCD does not.
In a CCD each pixel well collects unamplified photoelectrons until they
are read out, amplified {\it en masse}, and sent to an analog-to-digital 
converter (A/D) over a 10 ft. cable.  The cable and its connectors
are by far the weakest link in the instrument and can cause real
grief if not properly maintained.
In the Argos instrument the default
amplifier gain is set so that two electrons yield one ADU, and has a
noise equivalent of 8 electrons. This means that 64 or more electrons
must be collected before their stochastic noise and the amplifier noise
are equal.  We try to keep pixel counts below 40,000 to avoid the
onset of nonlinear behavior; the PMT instruments have a somewhat wider
dynamic range.
These are not major problems, but can hardly
be considered assets.  

\subsubsection{Short Exposure Times}

PMT detectors can count individual photon events, and these events
can be placed into accurate time bins as short as desired.
The CCD
detector must accumulate many photon events before they can be
measured, so the minimum time bins must be much longer.  The PMT
instruments work well in measuring lunar occultations where 1--2
millisecond time bins are required, and for optical pulsar measurements
where the time bins as short as 1 microsecond have been used
(Sanwal, Robinson, \& Stiening 1998). 
The smallest time bin available in the Argos instrument, 1s, is 
short enough for measuring pulsating white dwarf stars, but not short
enough for these other measurements.

\subsubsection{Proprietary Secrets}

Roper Scientific, the corporate owners of Princeton Instruments who
made the CCD camera, have a policy of not revealing technical details
of either their hardware or their control software; users are given
access to the camera's operation by a series of calls to a software
library they provide. The user must therefore treat the camera and
its controlling software as a black box, and can infer or measure
details of its operation (e.g. timings) only by doing experiments.
This makes trouble-shooting and time-critical program development
both slow and difficult. The Argos instrument, and its software,
are unlikely to evolve into more effective operation so long as
this policy is in place.

\section{SUMMARY: RESULTS TO DATE}

The Argos instrument was placed into regular operation (with a less
capable version of the software) in February 2002.  Until this writing
(April 2003) it has achieved the following results:

\begin{itemize}

\item{} Its increased sensitivity has allowed the identification of 36
variable white dwarf stars previously unknown (Mukadam \etal 2003a), 
more than doubling the
total number of these objects available for study.

\item{} The improved data quality, and the resulting improvement in
measuring the time of arrival of pulses from the DAV white dwarf
G\,117-B15A, has yielded the first measured $\dot{P}$ (Kepler \etal 2003) 
for the rate of change
of the 215\,s principle period 
(previous measurements had only set limits).
This is the first such measurement
for a star this cool (and this old), and opens the way to calibrating
by measurement, rather than by theory, the ages of the oldest stars
in our galaxy (e.g. Winget \etal 1987; Hansen \etal 2002).

\item{} The control software and a second CCD camera was used successfully
at Siding Spring Observatory on the 1m telescope in support of a Whole
Earth Telescope run in May, 2002. The F/8 Cassegrain focal position 
gave about the same plate scale, and the same size images, as the prime 
focus position on the 82-in (F/3.9) telescope.

\end{itemize}

The success of the instrument, despite its limitations, can best be
judged by its use: the PMT photometers have not been used on the 82-in
telescope since Argos was placed into regular operation.

\acknowledgements
We thank Dr. Frank Bash, Director of McDonald Observatory,
for providing the funds to purchase the CCD cameras, and the
Texas Advanced Research Program for operating funds under
grant ARP-0543. We thank Gary Hansen for designing the timer
card, Gordon Wesley and David Boyd for the mount design, and
Phillip McQueen for advice on the baffle design. We also thank
Antonio Kanaan for the use of his CCD data reduction routines, and
Darragh O'Donoghue for showing us that off-axis coma would not be
a problem.  We thank Denis Sullivan for his help in the commissioning
runs for both the CCD cameras, Argos and Cyclops.

\appendix
{\twocolumn

\section{APPENDIX}
\subsection{The Software Structure}

The ideal environment for a real-time program is to have a CPU
dedicated to its task, and to have complete control over all of its
operations. In an earlier era (before operating systems became
common) this was practical, but it meant the program had
to include many of the functions we now expect an operating system
to perform. Simple operating systems that supported only one user
and one task (e.g. MS-DOS) could be used despite some loss of
control. Multiuser, multitasking operating systems represent a
more significant challenge to a real-time program, but also offer
system capabilities that can help do the job, within limits.  The
Quilt 11 program is designed to function in this less predictable
environment, and seems to be successful at it.

A single CPU can only keep one process at a time in execution, but
it can be switched rapidly between several so they appear to be
running in parallel. The Linux kernel does this to support and
schedule many processes, most of which are dormant until needed
for the function they perform. The q11 program consists of five
such processes, three of them dormant most of the time. Figure 7
shows this software architecture as a flow chart.}

\subsection{Star Image Identification}

Before on-line extraction can begin, the user must first tell the program 
which star image in a field is the target star, and which other images it
should use as comparison stars. When the ``mark'' command is detected
the current image is copied to a separate image buffer, and the display 
process is directed there. New images will still arrive, but will not be 
shown until marking has been completed.

The process by which stars on a new image are identified with those 
initially marked is based on a simple assumption: that a star image may 
move around on the CCD chip, but its location with respect to other stars 
in the same field will not change (much). The identification algorithm 
therefore needs to remember a pattern of star images found in one image, 
so it can identify elements of the same pattern in a new one. Pattern
recognition is a famously difficult programming problem, but we are 
fortunate here on two counts: first, our patterns are very simple, and can
be reduced to a small list of x,y locations of points in a plane; second,
we are really interested in pattern re-recognition, which is a far
easier problem to solve.

To make this work, the program must first isolate the individual stars
in an image (described in the next section), and make a position list 
(plist) for each one, recording
its x,y location along with other useful values. The position is taken
to be the pixel with the highest count in it. From this position list,
the marking process can first identify which image the user indicates
with the mouse click, and can then make a reference list (rlist) of
distance and position angle (dpa) to other stars in the field for
later comparison.  

The identification algorithm finds a known star in a new image by
first deriving a set of dpa's for it (from the plist of image locations)
and then comparing them with the rlists saved from the marked stars.
In effect, it is asking of each star in a new field, ``Are you on my
list?''  Most often the answer is no -- the distances and position angles
to other field stars do not fit one of the saved patterns, except
by accident. An rlist can have up to 12 entries, and an accidental
match to more than 1 or 2 of them just doesn't happen, even in a crowded
field. Agreement with all of the rlist entries is common when a
true match is found, but not required: the majority wins.

One source of potential position error can arise if the rlist
values are taken from a single image and remain unchanged during a
run. An onset of bad seeing can cause fewer reference stars to be
recognized, and changing atmospheric refraction can affect the dpa
values in a systematic way. To avoid these problems, the dpa values
from each new image are saved (they were calculated to effect the
comparisons) and a new rlist is made for each marked star that is
identified. The rlist that participates in the comparison process
is actually a running average of the last 20 rlists encountered,
explaining the cryptic entry (``co-add dpa's'') in the flow chart.
To avoid contaminating the running average in case of clouds, the
rlist is not added in if the sky transparency becomes poor.

Even though the identification algorithm must examine every star in a
new image, it doesn't spend much time doing it. The richest field
we have encountered has about 230 stars detected in it, but the
identification process needs less than 100 milliseconds to do its job.
It uses correspondingly less time if the field is less crowded.

\subsection{On-Line Luminosity Extraction}

The arrival of a new image triggers a flurry of activity that results
in a new data point in the light curves of any marked stars.  First,
though, the star images must be isolated and located on the chip
before they can be identified. The process of extracting the
luminosity is melded in with the isolating procedure.

By analogy, we can think of the star images as luminosity mountains
rising as peaks above the plane of the sky. If we flood the plane,
and consider only the peaks that rise above flood level, then they are 
nicely isolated and can be treated individually.

We first approximate the sky level by finding the mean of all of the
pixel values in the image, avl, and then setting a cut level (flood
level) enough above that mean to avoid finding false peaks due to
sky noise:

\begin{equation}
cut = avl + 2.5 \times \sqrt{avl}
\end{equation}

We can now scan the image one pixel at a time, and determine for each
one if it is above the cut level, and therefore a part of some star
image, or below it, and part of the sky level. In the process we
look for successive pixels above the cut level, and contiguous with
those on previous scan lines, to define a luminosity ``clump''.
We consider
a clump as complete (and therefore isolated) when a scan in x finds
no more pixels to include. The plist entry for this peak then contains 
the location of the maximum (in x and y), its total luminosity above
the cut level, and the number of pixels summed.  

This procedure works well for peaks that are not too broad,
but can sometimes yield small false peaks near a large one when the
seeing is bad or the images are a bit out of focus. Akin to foothills
in our analogy, these ``skirt peaks'' may confuse the identification 
procedure. To avoid this, when the scan is complete, the plist is 
examined by a routine that identifies small peaks too close to big ones 
and merges them in. The resulting plist is then sorted by luminosity
so the brightest ones appear first; identification can now proceed.  

Once a peak has been identified with a user-marked star, its 
total luminosity (plus a small correction for the fraction of its 
luminosity below the cut level) becomes the next data point on its light 
curve, after a global sky value has been subtracted. This global sky 
value, found during the isolation scan, is the mean level of all the 
pixels below the cut level, and is really composed of sky photons, 
dark count, and a bias value set by the electronics to ensure only 
positive quantities are presented to the A/D converter.

The on-line extraction process was devised to provide the user with
light curves in real time, the same as with the PMT instrument, to
allow the data quality to be assessed. It was not intended to be
a final data reduction procedure. However,
cut-level extraction proves to work far better than originally expected,
and may evolve into a procedure that can rival the virtual aperture
extraction technique (O'Donoghue et al. 2000); alternatively, 
that procedure could
be incorporated into the q11 program as a user-selected alternative.

{\onecolumn
\begin{figure}[ht]
\figurenum{7}
\plotone{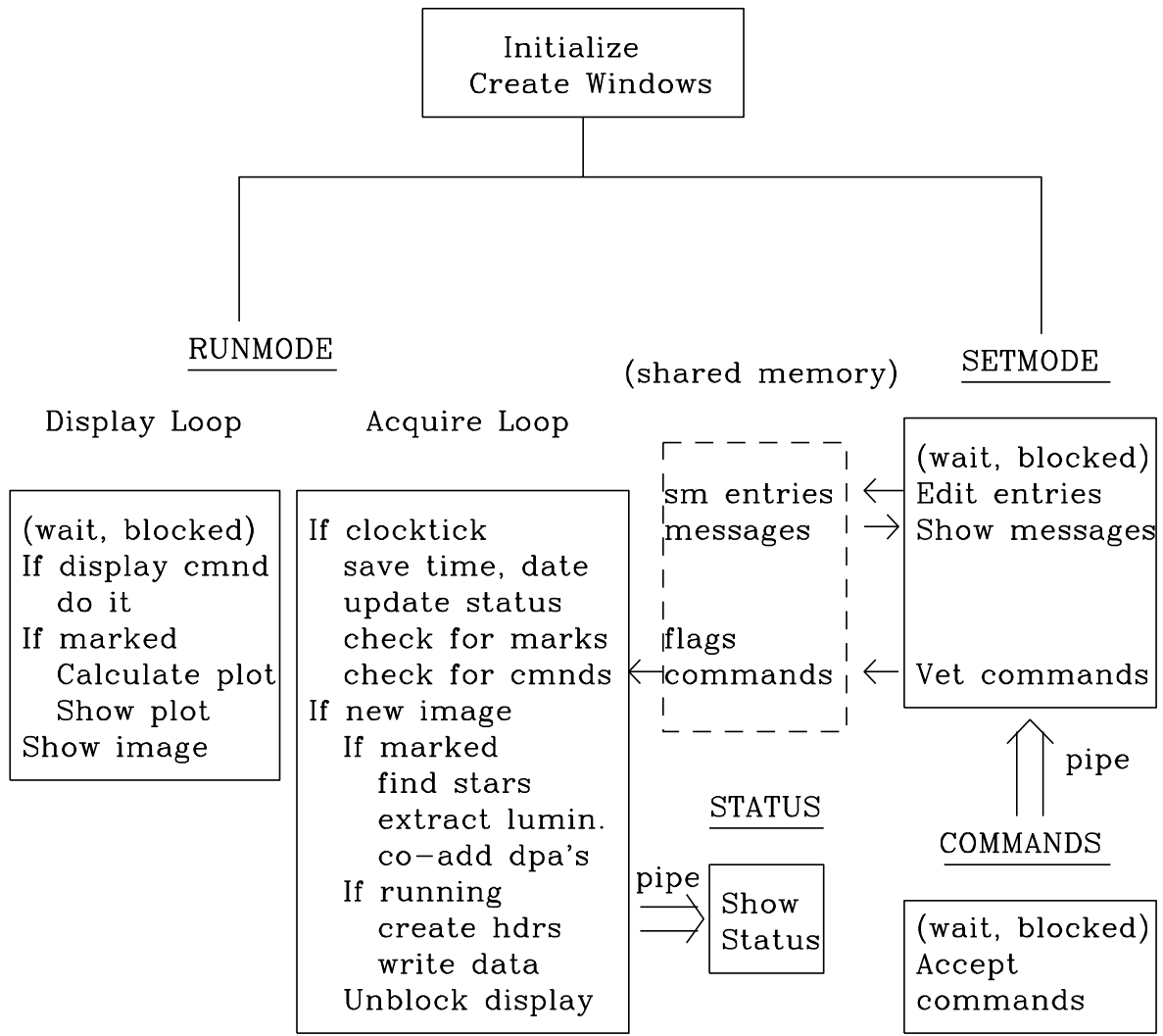}
\figcaption{Quilt 11 Architecture.  Five processes (solid boxes) can execute
in parallel, but four of them are blocked until needed.  The Acquire loop
polls the clock and the camera continuously, watching for the 1Hz clock
ticks or the arrival of a new image.}
\end{figure}
}
\twocolumn


\begin{thebibliography}{}

\bibitem[Hansen et al.(2002)]{2002ApJ...574L.155H} Hansen, B.~M.~S.~et al.,
2002, \apjl, 574, L155

\bibitem[Kanaan et al.(2000)]{2000BaltA...9..387K} Kanaan, A., O'Donoghue, 
D., Kleinman, S.~J., Krzesinski, J., Koester, D., \& Dreizler, S.\ 2000, 
Baltic Astronomy, 9, 387 

\bibitem[Kepler, et al. (2003)]{}Kepler, S. O.~et al.\
2003, ApJ, in preparation

\bibitem[Kepler, et al. (2000)]{2000ApJ...534L.185K} Kepler, S. O.,
Mukadam, A., Winget, D. E., Nather, R. E., Metcalfe, T. S., Reed, M. D.,
Kawaler, S. D., \& Bradley, P. A. 2000, \apjl, 534, L185

\bibitem[Kleinman, Nather \& Phillips (1996)]{1996PASP..108..356K} Kleinman,
S. J., Nather, R. E., \& Phillips, T. 1996, \pasp, 108, 356

\bibitem[Mukadam et al.(2003a)]{} Mukadam, A.~et al.\
2003a, ApJ, in preparation

\bibitem[Mukadam et al.(2003b)]{2003whdw.conf..227M} Mukadam, A.~et al.\
2003b, White Dwarfs, proceedings of the conference held at the Astronomical
Observatory of Capodimonte, Napoli,
Italy\footnote{http://www.na.astro.it/meetings/wd2002/wd.html}. To be published by
Kluwer.~(NATO Science Series II -- Mathematics, Physics and Chemistry,
Vol.~105, p.~227.), 227

\bibitem[Mullally et al.(2003)]{2003whdw.conf..337M} Mullally, F., Mukadam,
A., Winget, D.~E., Nather, R.~E., \& Kepler, S.~O.\ 2003, White Dwarfs,
proceedings of the conference held at the Astronomical Observatory of
Capodimonte, Napoli,
Italy.~To be published by
Kluwer.~(NATO Science Series II -- Mathematics, Physics and Chemistry,
Vol.~105, p.~337.), 337

\bibitem[Nather \& Warner(1971)]{1971MNRAS.152..209N} Nather, R.~E.~\&
Warner, B.\ 1971, \mnras, 152, 209

\bibitem[Nitta et al.(2000)]{2000BaltA...9...97N} Nitta, A., Kanaan, A., 
Kepler, S.~O., Koester, D., Montgomery, M.~H., \& Winget, D.~E.\ 2000, 
Baltic Astronomy, 9, 97 

\bibitem[Nitta et al.(1998)]{1998BaltA...7..203N} Nitta, A., Kepler, S.~O., 
Winget, D.~E., Koester, D., Krzesinski, J., Pajdosz, G., Jiang, X., \& 
Zola, S.\ 1998, Baltic Astronomy, 7, 203 

\bibitem[O'Donoghue et al.(2000)]{2000BaltA...9..375O} O'Donoghue, D.,
Kanaan, A., Kleinman, S.~J., Krzesinski, J., \& Pritchet, C.\ 2000, Baltic
Astronomy, 9, 375

\bibitem[Robinson et al.(1995)]{1995ApJ...438..908R} Robinson, E.~L.~et 
al.\ 1995, \apj, 438, 908 

\bibitem[Sanwal, Robinson, \& Stiening(1998)]{1998AAS...19311204S} Sanwal,
D., Robinson, E.~L., \& Stiening, R.~F.\ 1998, Bulletin of the American
Astronomical Society, 30, 1420

\bibitem[Winget et al.\ (1987)]{1987ApJ...315L..77W} Winget, D.\ E.,
Hansen, C.\ J., Liebert, J., van Horn, H.\ M., Fontaine, G., Nather, R.\
E., Kepler, S.\ O., \& Lamb, D.\ Q.\ 1987, \apjl, 315, L77

\bibitem[Winget et al.\ (2003)]{} Winget, D.\ E., Nather, R. E., Mukadam, A.,
Mullally, F., von Hippel, T., Cochran, W. D., Endl, M., Slaughter, D., Reaves,
D., Kepler, S. O., Kanaan, A., \& Sullivan, D. J.~ 2003, To be published in ASP Conf. Ser. 294,
Scientific Frontiers in Research on Extrasolar Planets, ed. D. Deming \& S. Seager.
 
\end{thebibliography}
\end{document}